# Screening by symmetry of long-range hydrodynamic interactions of polymers confined in sheets

**Tsvi Tlusty**

*Department of Physics of Complex Systems, Weizmann Institute of Science, Rehovot, Isreal*

*Email: tsvi.tlusty@weizmann.ac.il*

ABSTRACT: Hydrodynamic forces may significantly affect the motion of polymers. In sheet-like cavities, such as the cell's cytoplasm and micro-fluidic channels, the hydrodynamic forces are long-range. It is therefore expected that that hydrodynamic interactions will dominate the motion of polymers in sheets and will be manifested by Zimm-like scaling. Quite the opposite, we note here that although the hydrodynamic forces are long-range their overall effect on the motion of polymers vanishes due to the symmetry of the two-dimensional flow. As a result, the predicted scaling of experimental observables such as the diffusion coefficient or the rotational diffusion time is Rouse-like, in accord with recent experiments. The effective screening validates the use of the non-interacting blobs picture for polymers confined in a sheet.

## Introduction

When a polymer moves around inside a solution it drags the surrounding solvent and thus may mediate hydrodynamic forces between its monomers [1]. These forces are known to dramatically change the dynamic behavior of polymers as manifested by experimental observables such as the diffusion coefficient $D$, the rotational diffusion time $\tau_R$ or the dynamic scattering spectrum [2, 3]. Extensive studies of the hydrodynamic interactions have distinguished two asymptotic regimes: In the Rouse regime [4], the movement of each monomer is determined solely by the local forces. In this regime the scaling of $D$ and $\tau_R$ as the function of the polymer's molecular weight $M$ is $D \sim M^{-1}$ and $\tau_R \sim M^{2}$. At the other limit, the Zimm regime[5], the many-body hydrodynamic interactions between the monomers dominate their motion. The scaling in this regime changes to $D \sim M^{-1/2}$ and $\tau_R \sim M^{3/2}$, and may differ from the Rouse scaling by several orders-of magnitude. Hence, beside the theoretical interest, the question of whether hydrodynamic forces affect polymer dynamics has a vast practical importance in a variety of experimental setups and engineered devices where polymers are transported and manipulated [6, 7].

The present note discusses the effect of the hydrodynamic forces on polymer solutions that are confined in a sheet, a geometry where two of the dimensions are much larger than the third, the height $h$. This configuration is sometimes termed slit, crack or 'quasi two-dimensional' [7, 8]. The effect of the boundaries is expected to be significant when the sheet's height is smaller than the gyration radius of the polymer. Such geometry is typical to biopolymers confined in living cells or to macromolecules flowing through the cracks of a porous media. Flat channels are also the basic wiring of the rapidly developing micro-fabricated fluidic devices [6, 7] and their design therefore relies on the understanding of quasi-2D hydrodynamics.

A series of recent experimental and theoretical studies have demonstrated that colloidal beads suspended in a fluid sheet between two plates exhibit anomalously long-range hydrodynamic interactions [8-12]. Since the basic dynamic properties of polymers can be represented by a set of colloidal beads connected along a chain [1, 13], it is natural to expect that the same long-range forces also dominate the dynamics of a quasi-2D polymer solution and lead to Zimm-like scaling. In contrast, the simple conclusion of this note is that although the hydrodynamic forces are indeed long-range their overall effect vanishes by symmetry. As an immediate result, we predict Rouse-like scaling for polymers confined in flat channels or sheets, e.g. $D \sim M^{-1}$ and $\tau_R \sim M^2$, in accord with recent experiments [14] and simulation studies [15, 16].

In what follows, we first present a simplified derivation of the hydrodynamic forces in sheets. We then examine the impact of these forces on polymer dynamics and conclude that their net effect vanishes. Finally, we discuss the validity of this result to geometries where the boundaries are not parallel and its relevance to the blob picture[17, 18] of 2D polymer dynamics.

## Results and Discussion

**Quasi-2D hydrodynamics at low Reynolds number.** Polymers flow in a sheet typically at a very low Reynolds number and the hydrodynamics is therefore that of the viscous Stokes flow. It follows from the linearity of the Stokes flow that the velocities $\mathbf{u}_n$ of particles suspended in a viscous solution are linearly related to the forces acting on them $\mathbf{f}_n$ by the Oseen tensor **H** [19],

$$(1) \quad \mathbf{u}_n = \sum_m \mathbf{H}\left(\mathbf{r}_n - \mathbf{r}_m\right) \cdot \mathbf{f}_m \,,$$

where for the present 2D geometry **H** is a 2x2 tensor that depends on the distance between the particles $\mathbf{r}_n$-$\mathbf{r}_m$.

To examine the hydrodynamic forces in quasi-2D geometry we consider the flow-field around a coin-like disk of radius $a$ that moves horizontally at a velocity **U** in the middle of a sheet bounded by the $z = \pm h/2$ planes (figure 1). In principle, given the no-slip boundary conditions it is straightforward though rather cumbersome to derive the Oseen tensor from Stokes equations [20, 21]. Nonetheless, the equations can be further simplified by employing the lubrication approximation used in the Hele-Shaw cell [7, 22]: Since the height $h$ is much smaller than the horizontal dimensions, it restricts the profile of the velocity **v** to be approximately that of the parabolic Poiseuille flow,

$$(2) \quad \begin{aligned} \mathbf{v}\left(x,y,z\right) &= \mathbf{u}\left(x,y\right)\left(1-\left(\frac{2z}{h}\right)^2\right), \\ \mathbf{u}\left(x,y\right) &= -\frac{h^2}{8\eta}\nabla p\left(x,y\right), \end{aligned}$$

where $\eta$ is the viscosity of the host fluid and the mid-plane velocity $\mathbf{u}(x,y)$ is proportional to the gradient of the pressure $p$. Equation 2, termed sometimes the Darcy approximation, is equivalent to a 2D potential flow, $\mathbf{u}(x,y) = \nabla\phi$ with the effective potential $\phi = -h^2p/8\eta$. Accounting for incompressibility, $\nabla\cdot\mathbf{v} = 0$, one finds the Laplace equation, $\Delta\phi = 0$.

The flow around the disk is found by solving the Laplace equation with the boundary condition of zero mass-flux through the edge of the disk. This yields a potential and a velocity field of a two-dimensional mass-dipole of strength proportional to the area and velocity of the disk, $\mathbf{d} = a^2\mathbf{U}$,

$$(3) \qquad \begin{aligned} \phi &= -\frac{\mathbf{d}\cdot\hat{\mathbf{r}}}{r}, \\ \mathbf{u} &= \nabla\phi = \frac{1}{r^2}\left(2\left(\mathbf{d}\cdot\hat{\mathbf{r}}\right)\hat{\mathbf{r}} - \mathbf{d}\right). \end{aligned}$$

where $\hat{\mathbf{r}}$ is a unit vector in the direction of the position $\mathbf{r}$. Summing the momentum flux through the disk's edge we find the overall drag exerted by the host fluid, which is counterbalanced by the external force $\mathbf{f}$ required to move the disk, $\mathbf{f} \sim \eta\mathbf{d}/h$. And so, exerting a force $\mathbf{f}$ on a particle induces at $\mathbf{r}$ a velocity $\mathbf{u}$ given by (3), which in turn may drag other particles. This relation between the induced flow and the force determines the quasi-2D Oseen tensor (1),

$$(4) \qquad \mathbf{H}\left(\mathbf{r}\right) \sim \frac{h}{\eta r^2}\left(2\hat{\mathbf{r}}\otimes\hat{\mathbf{r}} - \mathbf{1}\right).$$

This dependence on distance and direction of the hydrodynamic interaction was derived in exact solutions for point forces ('stokeslets') confined between parallel plates [20, 21]. Nevertheless, the simplified derivation adds some insight by clarifying the relation to 2D potential flows. The dipolar flow (3) approximates well the far-field velocity induced by the monomers of a polymer in a sheet.

**The effect of hydrodynamic forces on polymers in a sheet.** To explore the basic dynamic properties of a polymer, its monomers are represented as a set of Brownian particles connected along a chain by elastic springs [13]. This classical picture is expressed by the generalized Langevin equation for the monomer positions $r_n$ [1],

$$(5) \qquad \mathbf{u}_n = \frac{d\mathbf{r}_n}{dt} = \sum_m \mathbf{H}\left(\mathbf{r}_n - \mathbf{r}_m\right)\cdot\left(-\frac{\partial E}{\partial \mathbf{r}_m} + \mathbf{f}_m(t)\right).$$

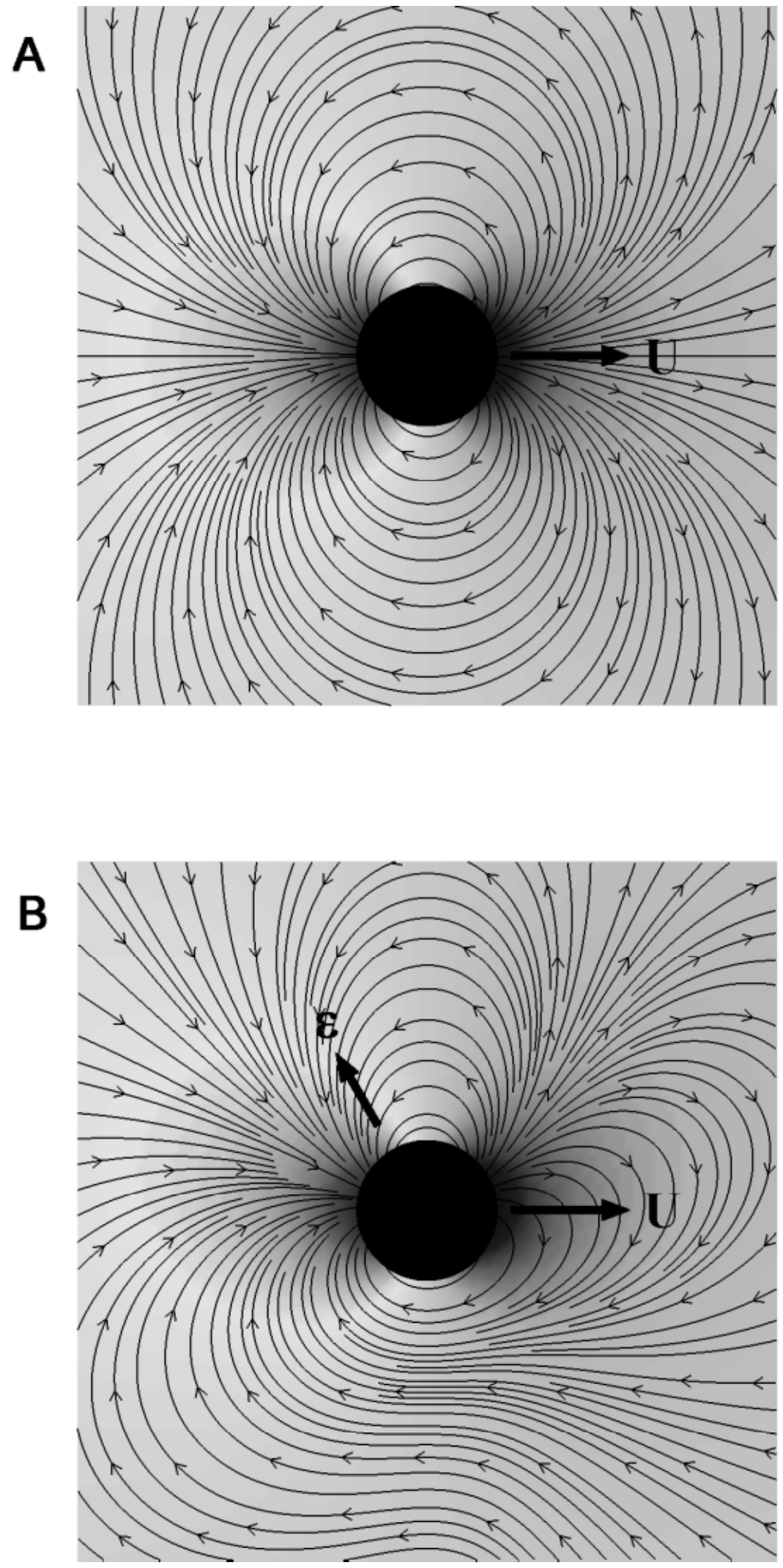

**Figure 1.** (a) Illustration of the flow field around a thin disk of radius $a$ moving in a uniform sheet at a velocity $\mathbf{U}$. The streamlines and the potential $\phi$ are that of a two-dimensional dipole of strength $\mathbf{d} = a^2\mathbf{U}$. It is evident that the orientational average of the velocity vanishes due to symmetry. (b) The same disk moving at the same velocity in a sheet of height that varies as $h(\mathbf{r}) = h_0\,(1 + \boldsymbol{\varepsilon}\cdot\mathbf{r})$, with a slope $|\boldsymbol{\varepsilon}|=1/(5a)$. Although the streamlines are significantly distorted the orientational average of the perturbed velocity ( 9) still vanishes.

This may be derived from (1) by noting that the overall force acting on the $m$-th monomer is the sum of the elastic force, which is the gradient of the elastic energy $E$, and the stochastic thermal force $\mathbf{f}_m$. Zimm has devised a way to simplify the analysis of the nonlinear equation (5) by 'pre-averaging' the Oseen tensor (4) over the equilibrium distribution of distances between the monomers $\mathbf{r}_{nm}= \mathbf{r}_n - \mathbf{r}_m$ [5],

$$(6) \quad \mathbf{H}\left(\mathbf{r}_{nm}\right) \rightarrow \left\langle \mathbf{H}\left(\mathbf{r}_{nm}\right)\right\rangle \sim \left( \text{[illegible]} \right) \left\langle 2\hat{\mathbf{r}}_{nm} \otimes \hat{\mathbf{r}}_{nm} - \mathbf{1} \right\rangle ,$$

where we used the fact that the ensemble averages over orientations and distances are independent.

The equilibrium average of the monomer-to-monomer separation scales as $\langle|\mathbf{r}_{nm}|^{-2}\rangle \sim |m\text{-}n|^{-2\nu}$, with Flory's exponent $\nu \approx 3/(D+2)$, where $\nu = 3/4$ in 2D and $\nu = 3/5$ in 3D. This algebraic decay of the average indicates strong hydrodynamic forces (Ref. 1 p. 93) and one might expect to observe the Zimm dynamics. However, due to the unusual symmetry of the Oseen tensor, its average over orientations vanishes, $\left\langle 2\hat{\mathbf{r}}_{nm} \otimes \hat{\mathbf{r}}_{nm} - \mathbf{1} \right\rangle = 0$. This immediately implies that the relevant hydrodynamic forces acting on a monomer are only due to the nearby monomers in the local 'blob'[17] of radius $h$ around the monomer [23, 24]. All the other, long-range hydrodynamic interactions may be disregarded. It follows directly that quasi-2D polymer solutions are always in the Rouse regime. We note that this conclusion relies on the accuracy of the pre-averaging approximation [25]. This result is in contrast to the situation in 3D polymer solutions where determination of the Rouse and Zimm regimes and the crossover between them is a matter of subtle modeling and observation [3].

The physical grounds that make the long-range hydrodynamic forces insignificant for the motion of quasi-2D polymers can be understood by re-examining the flow field in (3). The average of the Oseen tensor over all orientations (6) is actually the average over the dipole field that must vanish due to the symmetry of the dipole. As pointed out by the authors of references 8-10 this is very different from the situation in 3D Stokes flows that exhibit a momentum monopole. There, the pre-averaged contribution of the momentum monopole does not vanish and leads to the familiar Zimm scaling. In quasi-2D flows the monopole contribution decays exponentially because the confining plates absorb the momentum and the remaining leading term is the mass dipole.

**Screening of hydrodynamic forces in non-uniform sheets.** To explore the extent of validity of our conclusion that the overall effect of the hydrodynamic forces vanishes due to symmetry, we examine a geometry where the height of the fluid sheet is non-uniform (figure 1b). If the height $h$ changes moderately around an average $h_0$, we can linearly approximate it as $h(\mathbf{r}) \approx h_0\,(1 + \boldsymbol{\varepsilon}\cdot\mathbf{r})$, where the gradient is $\nabla h \approx h_0\boldsymbol{\varepsilon}$. This may describe a sharp wedge or a slit. The slope is moderate enough to keep the Darcy approximation (2) valid. However, since the velocity is proportional everywhere to $h^2$ and the flux to $h^3$ (integrating along the $z$-direction) the Laplace equation is modified into

$$(7)\quad \nabla\cdot\left(\left(h/h_0\right)^3\nabla\phi\right)=0,$$

with the effective potential $\phi = -h_0^2p/8\eta$. Substituting the linear dependence of $h$, equation 7 can be rewritten as $(1 + 3\boldsymbol{\varepsilon}\cdot\mathbf{r})\Delta\phi = -\,3\boldsymbol{\varepsilon}\cdot\nabla\phi$ .

Since the gradient $\boldsymbol{\varepsilon}$ is small we can expand the potential around the dipole flow of a uniform sheet $\phi_0$ (3), $\phi \approx \phi_0 + \phi_1$. The perturbation $\phi_1$ may therefore be found from $\Delta\phi_1 = -\,3\boldsymbol{\varepsilon}\cdot\nabla\phi_0$. Solution of the latter equation with the appropriate boundary condition on the disk gives

$$(8)\quad \phi_1=\frac{3}{4}\left(\mathbf{d}\cdot\boldsymbol{\varepsilon}-2\left(\mathbf{d}\cdot\hat{r}\right)\left(\boldsymbol{\varepsilon}\cdot\hat{r}\right)\right).$$

The corresponding perturbation $\mathbf{u}_1$ of the velocity, $\mathbf{u} = (h/h_0)^2\,\nabla\phi = \nabla\phi_0 + \mathbf{u}_1$ , is

$$(9)\quad \mathbf{u}_1=2\left(\boldsymbol{\varepsilon}\cdot\mathbf{r}\right)\nabla\phi_0+\nabla\phi_1=\frac{1}{2r}\left(11\left(\boldsymbol{\varepsilon}\cdot\hat{\mathbf{r}}\right)\left(\mathbf{d}\cdot\hat{\mathbf{r}}\right)\hat{\mathbf{r}}-\left(\boldsymbol{\varepsilon}\cdot\hat{\mathbf{r}}\right)\mathbf{d}-3\left(\mathbf{d}\cdot\hat{\mathbf{r}}\right)\boldsymbol{\varepsilon}\right)+o\left(\boldsymbol{\varepsilon}^2\right).$$

The resulting perturbation of the flow field is shown in figure 1b. Despite the significant distortion of the field, it is straightforward to verify that the average over all orientations of (9) vanishes, $\langle \mathbf{u}_1 \rangle=0$. Thus, up to first order in the gradient ε, the pre-averaged hydrodynamic interaction vanishes in moderately non-uniform fluid sheets. This implies that the hydrodynamic interactions of a polymer are effectively screened also in non-uniform sheets if the height $h$ does not change much over the polymer's radius of gyration $R_g$, $|\boldsymbol{\varepsilon}|R_g << 1$.

## Conclusions

The conclusion that the hydrodynamic forces are screened due to symmetry relies only on the soundness of the Darcy approximation. It will therefore remain valid even if the sheet is curved, for example into a spherical or cylindrical shell, as long as the height does not vary too steeply.

The dipolar flow described here is not unique to polymer solutions in sheets. It was observed in a number of two-dimensional configurations [26-28]. Screening of long-range hydrodynamic forces was noted in electrophoresis [27] and for Brownian particles in a thin film [28], where the momentum flux is dissipated by a phenomenological friction force. It can be expected to observe similar screening effects in other configurations whenever a mechanism that dissipates momentum is present.

Finally, we remark that the present effective screening justifies the use of the blob picture[17] to derive the scaling of the diffusion coefficient and other related observables of polymer solutions in thin sheets [23, 24]. This is because the blob scaling relies on the fact that blobs do not exert hydrodynamic forces on each other due to the strong screening. The total friction of the polymer is therefore proportional to the number of blobs $N_b$ times the friction of a single blob, which by itself is proportional to the blob size $h$. Taking into account the dependence of $N_b$ on $h$ and on the polymer's mass $M$, the blob scaling of the diffusion coefficient is Rouse-like, $D \sim (N_b h)^{-1} \sim h^{1/\nu-1} M^{-1}$. This scaling of the diffusion coefficient is well-known for polymers confined in a tube[17], where the hydrodynamic forces decay exponentially and the polymer therefore exhibits Rouse-like dynamics[18, 24]. The present study suggests that the same scaling holds also for polymers confined in a thin sheet, although for very different reasons. Unlike the flow in the tube, the long-range hydrodynamic forces between monomers of a polymer confined in a sheet are not screened, yet their overall effect vanishes due to symmetry of the flow field.

The blob scaling agrees with recent diffusivity measurements of double-stranded DNA molecules confined in a slit-like channel [14], where the blob scaling is presented in the equivalent normalized form $D/D_{bulk} \sim (R_g/h)^{1-1/\nu}$. Nevertheless, the current measurement has reached only moderate values of the confinement parameter, $R_g/h \sim 0.25$-$0.9$. In the high confinement regime of thinner slits and longer polymers, the blob scaling is supported only by simulation studies [15, 16] and is yet to be experimentally tested. We expect the $h^{1/\nu-1}$ dependence of the blob scaling to break down for ultra-thin sheets [29] whose height is of the order of a few persistence lengths. The $M^{-1}$ dependence merely manifests the locality of the forces and is therefore expected to remain valid even for very thin sheets [30].